\documentclass[letterpaper,10pt,twoside]{article}

\usepackage[utf8]{inputenc}  
\usepackage[english]{babel} 
\usepackage[english=american]{csquotes}
\MakeOuterQuote{"}

\usepackage{lmodern}
\usepackage{amssymb,amsthm,mathtools} 
\usepackage{hyperref} 
\usepackage{geometry} 

\newcommand{\im}{\mathrm{i}}

\newcommand{\R}{\mathbb{R}}
\newcommand{\C}{\mathbb{C}}
\newcommand{\defeq}{\coloneqq}
\newcommand{\tens}{\otimes}
\DeclareMathOperator{\id}{id}
\newcommand{\xd}{\mathrm{d}}
\newcommand{\cH}{\mathcal{H}}

\newcommand{\coh}{\mathsf{K}}
\newcommand{\ncoh}{\mathsf{k}}
\newcommand{\ls}{\ell}
\newcommand{\ms}{\mathsf{m}}
\newcommand{\oR}{\overline{R}}
\newcommand{\tord}{\mathbf{T}}
\newcommand{\dop}{\mathsf{D}}

\newcommand{\Lp}{\mathrm{p}}
\newcommand{\Le}{\mathrm{e}}
\newcommand{\Lin}{\mathrm{in}}
\newcommand{\Lout}{\mathrm{out}}
\newcommand{\Lcin}{{\overline{\mathrm{in}}}}
\newcommand{\Lcout}{{\overline{\mathrm{out}}}}
\newcommand{\Lx}{{\mathrm{x}}}
\newcommand{\Li}{{\mathrm{i}}}
\newcommand{\Lcx}{{\overline{\mathrm{x}}}}
\newcommand{\Lci}{{\overline{\mathrm{i}}}}
\newcommand{\Lsi}{{\mathrm{sh}}}
\newcommand{\Lco}{{\mathrm{ch}}}

\allowdisplaybreaks

\begin{document}


\begin{titlepage}
\title{\textbf{Scattering of Evanescent Particles}}
\author{%
  Robert Oeckl\footnote{email: robert@matmor.unam.mx}\\
  Centro de Ciencias Matemáticas,\\
  Universidad Nacional Autónoma de México,\\
  C.P.~58190, Morelia, Michoacán, Mexico}
\date{UNAM-CCM-2021-2\\ 16 May 2021\\ 4 November 2022 (v2)}

\maketitle

\vspace{\stretch{1}}

\begin{abstract}

Massive Klein-Gordon theory is quantized on the timelike hypercylinder in Minkowski space. Crucially, not only the propagating, but also the evanescent sector of phase space is included, laying in this way foundations for a quantum scattering theory of fields at finite distance. To achieve this, the novel $\alpha$-Kähler quantization scheme is employed in the framework of general boundary quantum field theory. A potential quantization ambiguity is fixed by stringent requirements, leading to a unitary radial evolution. Formulas for building scattering amplitudes and correlation functions are exhibited. A novel LSZ formula is derived, applicable to scattering at finite distance.

\end{abstract}

\vspace{\stretch{1}}
\end{titlepage}


\section{Introduction}

Evanescent waves play an important role in many electromagnetic phenomena. However, they are not part of the Hilbert space of standard quantum field theory in Minkowski space. This Hilbert space arises from the quantization of the free field in a neighborhood of a spacelike hypersurface. Since evanescent waves in vacuum are unbounded at spatial infinity, they do not form part of the phase space of well-behaved solutions that can be quantized. Instead, only propagating waves are quantized. The main predictive instrument of quantum field theory (at least in high energy physics) is the S-matrix which is based on the idealization that the observer is infinitely removed from the interaction region. It is then physically correct that the observer should not see any evanescent waves. Phenomena involving evanescent waves are thus described by standard quantum field theory only indirectly, through their imprints on propagating quanta reaching the observer far away.

At finite distance from the interaction region, on the other hand, evanescent waves do appear, and measurement devices will detect them. If we want to describe such measurements directly, we need a quantum description of the degrees of freedom of the evanescent field. One approach to do so was pioneered by Carniglia and Mandel in the seminal paper \cite{CarMan:quantevanemw}. It consists of introducing a medium with a refractive index larger than one in a region of space (such as a half-space) and then quantizing the corresponding phase space of classical solutions by the usual methods of free field theory. The solutions will consist of propagating waves, but also contain evanescent contributions near the interface between the medium and the vacuum. While this approach certainly has its merits, it is inherently limited in various respects. Firstly, it does not provide a quantum description of the evanescent field alone, i.e., of “evanescent particles”. Secondly, it is applicable only in a very particular class of physical situations where a classical background medium is a good approximation. In particular, even if a medium is involved in the phenomenon to be modeled, this approach precludes a possibly desirable description of the medium itself in terms of more fundamental degrees of freedom.

A direct quantization of evanescent waves has been achieved only recently \cite{CoOe:evanescent}. This has required a novel quantization scheme \cite{CoOe:locgenvac}, extending standard canonical quantization to the realm applicable for evanescent waves \cite{CoOe:vaclag}. Underlying this is the framework of general boundary quantum field theory (GBQFT) \cite{Oe:boundary,Oe:gbqft,Oe:holomorphic,Oe:feynobs} which generalizes the standard methods of quantum field theory (QFT) to deal with the description of physics in finite pieces of spacetime and on general (not necessarily spacelike) hypersurfaces in a coherent way.

In the standard approach to quantum field theory, only one Hilbert space of states is considered. This Hilbert space arises from the quantization of the phase space of classical solutions in a neighborhood of a spacelike hypersurface. By the Cauchy property, the phase spaces for different spacelike hypersurfaces in Minkowski space can all be identified with a phase space of global solutions, leading to “the” Hilbert space of states upon quantization.
In particular, this phase space (and the resulting Hilbert space) exclude evanescent waves, as these are unbounded when extended over all of space.
In contrast, in GBQFT it is possible (and necessary if results beyond the reach of standard QFT are to be obtained) to quantize the phase spaces of solutions in neighborhoods of general hypersurfaces. Evanescent waves do appear in the phase space of a timelike hypersurface if they are bounded in the spatial directions along the hypersurface.

The first quantization of the phase space on a timelike hypersurface in field theory was performed in \cite{Oe:timelike}. This was the quantization of Klein-Gordon theory on a plane in space, extended over all of time (the timelike hyperplane). This quantization, while successful for the propagating sector, excluded the evanescent sector. This was due to what appeared at the time to be technical issues with the Schrödinger representation. The underlying cause was fully understood only much more recently \cite{CoOe:vaclag}. It turns out that the physically correct vacuum boundary condition for the evanescent sector corresponds to a real Lagrangian subspace of the phase space. This is in contrast to the propagating sector and to the situation on spacelike hypersurfaces, where the vacuum boundary condition corresponds to a positive-definite (and necessarily complex) Lagrangian subspace of the phase space \cite{BiDa:qftcurved}. It is only in the latter case that the standard quantization prescription (to which we shall refer as \emph{Kähler quantization}) makes sense. Thus, a generalization of Kähler quantization had to be invented in order to address this novel situation \cite{CoOe:locgenvac}. Armed with this, it was possible to perform the quantization also of the evanescent sector of the phase space for the timelike hyperplane \cite{CoOe:evanescent}.

In the present work we lay the groundwork for quantum scattering theory at finite distance for the evanescent field. To this end we again use Klein-Gordon theory and adopt the timelike hypercylinder setting pioneered in \cite{Oe:kgtl}. We consider a sphere in space, marking the boundary of the interaction region of interest in the interior. On this boundary (which is a hypercylinder in spacetime) we describe the quantum degrees of freedom not only of incoming and outgoing propagating particles (as already done in \cite{Oe:kgtl}), but crucially also of the evanescent particles. The construction of the Hilbert space and associated structures is performed in Section~\ref{sec:quanthypcyl}. Particle states characterized by energy and angular momentum quantum numbers are introduced in Section~\ref{sec:particles}, both for the propagating and evanescent sectors. Coherent states and the action of observables are considered in Section~\ref{sec:cohsobs}. The radial evolution of states is rather simple as we shall see in Section~\ref{sec:radevol}. In Section~\ref{sec:ampl} we discuss scattering amplitudes for the free theory.  Correlation functions which are the building blocks of an interacting theory are the subject of Section~\ref{sec:correlation}. In Section~\ref{sec:lsz} we derive an LSZ formula for scattering at finite distance. We close with some final remarks in Section~\ref{sec:conclude}.


\section{Vacua and Hilbert spaces on the timelike hypercylinder}
\label{sec:quanthypcyl}

We consider in the following timelike hypersurfaces of the form $\R\times S^2_R$, where $S^2_R$ denotes the two-sphere in space of radius $R$, centered at the origin. The factor $\R$ represents the extension of the sphere over all of time. We refer to $\R\times S^2_R$ as the \emph{hypercylinder} of radius $R$. We use a cartesian time coordinate $t$ and spherical coordinates $(r,\phi,\theta)$ in space. We also denote the angular coordinates $(\phi,\theta)$ collectively by $\Omega$. $\xd\Omega$ denotes the standard measure on the $2$-sphere of unit radius.
We denote by $L_R$ the space of germs of solutions of the Klein-Gordon in a neighborhood of the hypercylinder of radius $R$. In the massless case, $L_R$ consists of propagating solutions only. Their quantization was carried out successfully in \cite{Oe:kgtl} using the Schrödinger representation. An equivalent quantization in terms of the holomorphic representation was presented in \cite{Oe:holomorphic}. In the massive case, $L_R$ splits into a direct sum $L_R^{\Lp}\oplus L_R^{\Le}$ of the space $L_R^{\Lp}$ of propagating solutions and the space $L_R^{\Le}$ of evanescent solutions. We mostly follow here the conventions and notations in \cite{CoOe:vaclag} which also provides some of the ingredients for the quantization of the evanescent waves.

We parametrize elements of $L_R^{\Lp,\C}$, i.e., complexified \emph{propagating solutions} ($E>m$) as follows,
\begin{multline}
 \phi(t,r,\Omega)=\int_{m}^{\infty}\xd E\, \frac{p}{4\pi} \sum_{\ls,\ms}
 \left(\left(\phi_{E,\ls,\ms}^\Lout h_{\ls}(p r) + \phi_{E,\ls,\ms}^\Lin \overline{h_{\ls}(pr)}\right) e^{-\im E t} Y_{\ls}^{\ms}(\Omega) \right.\\
 \left. + \left(\phi_{E,\ls,\ms}^\Lcout \overline{h_{\ls}(p r)} + \phi_{E,\ls,\ms}^\Lcin h_{\ls}(pr)\right) e^{\im E t} Y_{\ls}^{-\ms}(\Omega)\right) .
\label{eq:pparam}
\end{multline}
Here $Y_{\ls}^{\ms}$ denote the spherical harmonics and $p\defeq\sqrt{|E^2-m^2|}$. Also, $h_{\ls}=j_{\ls}+\im n_{\ls}$,
where $j_{\ls}$ and $n_{\ls}$ are the spherical Bessel functions of the first and second kind respectively. The space of real solutions is,
\begin{equation}
  L_R^\Lp=\{\phi\in L_R^{\Lp,\C} : \phi_{E,\ls,\ms}^\Lcout=\overline{\phi_{E,\ls,\ms}^\Lout},\; \phi_{E,\ls,\ms}^\Lcin=\overline{\phi_{E,\ls,\ms}^\Lin}\} .
\end{equation}
The oscillatory behavior of the solutions can be appreciated from asymptotic expansions of the spherical Bessel functions for large radius $r$ \cite[10.49(i)]{NIST:DLMF}:
\begin{equation}
    h_{\ls}(pr)=e^{\im p r}\sum_{k=0}^{\ls} \frac{\im^{k-\ls-1} (\ls+k)!}{2^k k! (\ls-k)! (pr)^{k+1}},
    \qquad \overline{h_{\ls}(pr)}=e^{-\im p r}\sum_{k=0}^{\ls} \frac{(-\im)^{k-\ls-1} (\ls+k)!}{2^k k! (\ls-k)! (pr)^{k+1}} .
\end{equation}
We can describe solutions as \emph{incoming} and as \emph{outgoing} (with respect to the interior of the hypercylinder) as already suggested by the notation for the parametrization:
\begin{align}
  L_R^{\Lin,\C} & =\{\phi\in L_R^{\Lp,\C} : \phi_{E,\ls,\ms}^\Lout=0,\;  \phi_{E,\ls,\ms}^\Lcout=0\},
  \label{eq:inmodes} \\
  L_R^{\Lout,\C} & =\{\phi\in L_R^{\Lp,\C} : \phi_{E,\ls,\ms}^\Lin=0,\;  \phi_{E,\ls,\ms}^\Lcin=0\} .
  \label{eq:outmodes}
\end{align}
We have the direct sum decomposition $L_R^{\Lp,\C}=L_R^{\Lin,\C}\oplus L_R^{\Lout,\C}$.
Solutions that are regular in the interior $M$ of the hypercylinder are those where the spherical Bessel functions $h_{\ls}$ and $\overline{h_{\ls}}$ combine to the spherical Bessel functions of the first kind $j_{\ls}$. We denote the subspace of these solutions by $L_M^{\Lp}\subseteq L_R^{\Lp}$. The complexified version of this is,
\begin{equation}
  L_M^{\Lp,\C}
  =\{\phi\in L_R^{\Lp,\C} : \phi_{E,\ls,\ms}^\Lin=\phi_{E,\ls,\ms}^\Lout,\; \phi_{E,\ls,\ms}^\Lcin=\phi_{E,\ls,\ms}^\Lcout \} .
  \label{eq:propint}
\end{equation}

We parametrize elements of $L_R^{\Le,\C}$, i.e., complexified \emph{evanescent solutions} ($E<m$) as follows,
\begin{multline}
 \phi(t,r,\Omega)=\int_{0}^{m}\xd E\, \frac{p}{2\pi^2}
  \sum_{\ls,\ms} \left(\left(\phi_{E,\ls,\ms}^\Lx k_{\ls}(p r)
+ \phi_{E,\ls,\ms}^\Li \tilde{k}_{\ls}(p r) \right) e^{-\im E t} Y_{\ls}^{\ms}(\Omega) \right. \\
+ \left. \left(\phi_{E,\ls,\ms}^\Lcx k_{\ls}(p r)
+ \phi_{E,\ls,\ms}^\Lci \tilde{k}_{\ls}(p r) \right) e^{\im E t} Y_{\ls}^{-\ms}(\Omega) \right) .
\label{eq:eparam}
\end{multline}
Here, $p\defeq\sqrt{|E^2-m^2|}$. Also, $k_{\ls}(z)=-\im^{\ls}\pi h_{\ls}(\im z)/2$ and $\tilde{k}_{\ls}(z)=k_{\ls}(-z)$ are modified spherical Bessel functions that are real for $z\in\R$. Note also that $a_{\ls}(z)\defeq (-1)^{\ls}k_{\ls}(z)+\tilde{k}_{\ls}(z)=\im^{-\ls+2}\pi j_{\ls}(\im z)$ is entire, i.e., in particular regular at the origin. We also define $\tilde{a}_{\ls}(z)\defeq k_{\ls}(z) - (-1)^{\ls} \tilde{k}_{\ls}(z)=\im^{\ls-1}\pi n_{\ls}(\im z)$. $k_{\ls}$, $\tilde{k}_{\ls}$ and $\tilde{a}_{\ls}$ have singularities at the origin, and $k_{\ls}(z)$ decays exponentially for increasing $z$. As long as $z$ is not small, $\tilde{k}_{\ls}(z)$ decays exponentially for decreasing $z$. The space of real solutions is,
\begin{equation}
  L_R^\Le=\{\phi\in L_R^{\Le,\C} : \phi_{E,\ls,\ms}^\Lcx=\overline{\phi_{E,\ls,\ms}^\Lx},\; \phi_{E,\ls,\ms}^\Lci=\overline{\phi_{E,\ls,\ms}^\Li}\} .
\end{equation}
Asymptotic expansions of the modified spherical Bessel functions show their behavior for large radius $r$ \cite[10.49(ii)]{NIST:DLMF}:
\begin{equation}
    k_{\ls}(pr)= e^{-p r}\sum_{k=0}^{\ls} \frac{\pi\, (\ls+k)!}{2^{k+1} k! (\ls-k)! (pr)^{k+1}},
    \qquad \tilde{k}_{\ls}(pr)= e^{p r}\sum_{k=0}^{\ls} \frac{\pi\, (-1)^{k+1} (\ls+k)!}{2^{k+1} k! (\ls-k)! (pr)^{k+1}} .
    \label{eq:asymptk}
\end{equation}
The evanescent solutions that are regular in the interior $M$ of the hypercylinder are the ones where the modified spherical Bessel functions $k_{\ls}$ and $\tilde{k}_{\ls}$ combine to $a_{\ls}$. We denote the subspace of those solutions by $L_M^{\Le}\subseteq L_R^{\Le}$. Explicitly,
\begin{equation}
  L_{M}^{\Le,\C}=\{\phi\in L_R^{\Le,\C} : \phi_{E,\ls,\ms}^{\Li}=(-1)^{\ls}\phi_{E,\ls,\ms}^{\Lx},\; \phi_{E,\ls,\ms}^{\Lci}=(-1)^{\ls}\phi_{E,\ls,\ms}^{\Lcx}\} .
  \label{eq:evanint}
\end{equation}

Important ingredients for the description of the classical dynamics and its quantization are the symplectic potential and symplectic form \cite{Woo:geomquant}. The \emph{symplectic potential} on the space of solutions $L_R$ associated to the hypercylinder of radius $R$, oriented as the boundary of the corresponding solid hypercylinder, is the bilinear form
\begin{equation}
 \theta_R(\phi,\xi) \defeq (\theta_{R})_{\phi}(\xi) = -R^2 \int \xd t\,\xd\Omega\,
 \xi(t,R,\Omega) \partial_r \phi(t,R,\Omega) .
\end{equation}
The \emph{symplectic form} arises as the antisymmetric part of the symplectic potential,
\begin{equation}
  \omega_R(\phi,\xi) = \frac{1}{2} \theta_R(\phi,\xi) - \frac{1}{2} \theta_R(\xi,\phi)
  =-\frac{R^2}{2}\int \xd t\,\xd\Omega\,
 \left(\xi(t,R,\Omega) \partial_r \phi(t,R,\Omega)- \phi(t,R,\Omega) \partial_r \xi(t,R,\Omega)\right)
\end{equation}
In the present parametrization and separating propagating and evanescent components we obtain,
\begin{align}
  \theta_R^{\Lp}(\phi,\xi)
  & = -R^2 \int_{m}^\infty\xd E\frac{p^3}{8\pi}\sum_{\ls,\ms}
  \left(\left(\phi_{E,\ls,\ms}^\Lout\xi_{E,\ls,\ms}^\Lcout
  + \phi_{E,\ls,\ms}^\Lcin\xi_{E,\ls,\ms}^\Lin\right) h_{\ls}'(p R) \overline{h_{\ls}(p R)} \right. \nonumber \\
  & \qquad +\left(\phi_{E,\ls,\ms}^\Lcout \xi_{E,\ls,\ms}^\Lout
  + \phi_{E,\ls,\ms}^\Lin \xi_{E,\ls,\ms}^\Lcin\right) \overline{h_{\ls}'(p R)} h_{\ls}(p R)
  \nonumber \\
  & \qquad +\left(\phi_{E,\ls,\ms}^\Lout \xi_{E,\ls,\ms}^\Lcin
  + \phi_{E,\ls,\ms}^\Lcin \xi_{E,\ls,\ms}^\Lout\right) h_{\ls}'(p R) h_{\ls}(p R) \nonumber \\
  & \qquad \left. +\left(\phi_{E,\ls,\ms}^\Lcout \xi_{E,\ls,\ms}^\Lin
  +\phi_{E,\ls,\ms}^\Lin \xi_{E,\ls,\ms}^\Lcout\right)
  \overline{h_{\ls}'(p R)} \overline{h_{\ls}(p R)}\right), \\
  \theta_R^{\Le}(\phi,\xi) & = -R^2 \int_{0}^{m}\xd E\frac{p^3}{2\pi^3}\sum_{\ls,\ms} \left(\left(\phi_{E,\ls,\ms}^\Lx \xi_{E,\ls,\ms}^\Lcx + \phi_{E,\ls,\ms}^\Lcx \xi_{E,\ls,\ms}^\Lx\right) k_{\ls}'(p R) k_{\ls}(p R) \right. \nonumber \\
  & \qquad +\left(\phi_{E,\ls,\ms}^\Li \xi_{E,\ls,\ms}^\Lci+\phi_{E,\ls,\ms}^\Lci \xi_{E,\ls,\ms}^\Li\right) \tilde{k}_{\ls}'(p R) \tilde{k}_{\ls}(p R) \nonumber \\
  & \qquad +\left(\phi_{E,\ls,\ms}^\Lx \xi_{E,\ls,\ms}^\Lci
  +\phi_{E,\ls,\ms}^\Lcx \xi_{E,\ls,\ms}^\Li\right) k_{\ls}'(p R) \tilde{k}_{\ls}(p R) \nonumber \\
  & \qquad \left .
    +\left(\phi_{E,\ls,\ms}^\Li \xi_{E,\ls,\ms}^\Lcx
    +\phi_{E,\ls,\ms}^\Lci \xi_{E,\ls,\ms}^\Lx
    \right) \tilde{k}_{\ls}'(p R) k_{\ls}(p R) \right) , \\
  \omega_R^{\Lp}(\phi,\xi)
  & = \int_{m}^\infty\xd E\frac{\im p}{8\pi}\sum_{\ls,\ms}
  \left(\xi_{E,\ls,\ms}^\Lout \phi_{E,\ls,\ms}^\Lcout
  +\xi_{E,\ls,\ms}^\Lcin \phi_{E,\ls,\ms}^\Lin
  -\xi_{E,\ls,\ms}^\Lcout \phi_{E,\ls,\ms}^\Lout
  -\xi_{E,\ls,\ms}^\Lin \phi_{E,\ls,\ms}^\Lcin\right) , \\
  \omega_R^{\Le}(\phi,\xi)
  & = \int_0^{m}\xd E\frac{p}{8\pi}\sum_{\ls,\ms}
  \left(\xi_{E,\ls,\ms}^\Lx\phi_{E,\ls,\ms}^\Lci
  + \xi_{E,\ls,\ms}^\Lcx\phi_{E,\ls,\ms}^\Li
  -\xi_{E,\ls,\ms}^\Li\phi_{E,\ls,\ms}^\Lcx
  -\xi_{E,\ls,\ms}^\Lci\phi_{E,\ls,\ms}^\Lx\right) .
\end{align}

For the propagating solutions quantization proceeds in analogy to standard Kähler quantization, except for the difference that we are on a timelike hypersurface instead of a spacelike one. This was carried out first in \cite{Oe:kgtl} using the Schrödinger representation and later in \cite{Oe:holomorphic} using the holomorphic representation. We briefly recall this from the perspective developed in \cite{CoOe:vaclag}. Thus, the vacuum in the exterior $X$ of the solid hypercylinder $M$ with radius $R$ is determined by a Lagrangian subspace $L_X^{\Lp,\C}$ of the space of germs of complexified propagating solutions $L_{\partial X}^{\Lp,\C}$ at the boundary $\partial X$ of $X$. Here, $L_{\partial X}^{\C}=L_{\oR}^{\C}$, where $\oR$ refers to the hypercylinder at radius $R$, but with opposite orientation. Recall that the symplectic form changes sign under orientation change \cite{Oe:holomorphic}. That is, $\omega_{\oR}=-\omega_R$. But note that the space of germs of solutions itself is the same, $L_{\oR}=L_R$.

The Wick rotated boundary condition at infinite radius amounts to a restriction to the $h_{\ls}$ modes,
\begin{equation}
  L_{X}^{\Lp,\C}=\{\phi\in L_R^{\Lp} : \phi_{E,\ls,\ms}^\Lin=0, \phi_{E,\ls,\ms}^\Lcout=0\} .
\end{equation}
This makes the sesquilinear form
\begin{equation}
  (\phi,\xi)_{\oR}=4\im\omega_{\oR}(\overline{\phi},\xi)
  \label{eq:stdip}
\end{equation}
positive definite on its restriction to $L_{X}^{\Lp,\C}$,
\begin{equation}
  (\phi,\xi)_{\oR}^{\Lp} =
  \int_{m}^\infty\xd E\frac{p}{2\pi}\sum_{\ls,\ms}
  \left(\xi_{E,\ls,\ms}^\Lout\overline{\phi_{E,\ls,\ms}^\Lout}
  +\xi_{E,\ls,\ms}^\Lcin\overline{\phi_{E,\ls,\ms}^\Lcin}
  -\xi_{E,\ls,\ms}^\Lcout\overline{\phi_{E,\ls,\ms}^\Lcout}
  -\xi_{E,\ls,\ms}^\Lin\overline{\phi_{E,\ls,\ms}^\Lin}\right) .
  \label{eq:propip}
\end{equation}
$L_{X}^{\Lp,\C}$ thus defines a Kähler polarization which encodes (part of) the standard vacuum in Minkowski space \cite{CoOe:vaclag}. Second quantization of this space with the inner product (\ref{eq:propip}) determines the Hilbert space $\cH_{R}^{\Lp}=\cH_{\partial M}^{\Lp}$ of states of the propagating degrees of freedom on (the exterior of) the hypercylinder of radius $R$. We decompose $L_R^{\Lp,\C}=L_{X}^{\Lp,\C} \oplus \overline{L_{X}^{\Lp,\C}}$ with notation $\phi=\phi^+ + \phi^-$ and consider the complex bilinear form
\begin{equation}
  \{\phi,\xi\}_{\oR}^{\Lp}=4\im\omega_{\oR}^{\Lp}(\phi^-,\xi^+)
  =\int_{m}^\infty\xd E\frac{p}{2\pi}\sum_{\ls,\ms}
  \left(\xi_{E,\ls,\ms}^\Lout\phi_{E,\ls,\ms}^\Lcout
  + \xi_{E,\ls,\ms}^\Lcin\phi_{E,\ls,\ms}^\Lin\right) .
  \label{eq:pbilin}
\end{equation}
This is positive definite on $L_R^{\Lp}$, the space of real propagating solutions. It defines the commutation relations between creation and annihilation operators, which are, labeled by elements $\phi,\eta\in L_R^{\Lp}$,
\begin{equation}
  [a_\eta,a_\phi^\dagger]=\{\phi,\eta\}_{\oR}^{\Lp} .
  \label{eq:pcom}
\end{equation}

For evanescent modes the physical vacuum in the exterior of the hypercylinder is given by a (non-Wick-rotated) decaying boundary condition \cite{CoOe:vaclag}. This is encoded in the Lagrangian subspace $L_{X}^{\Le,\C}\subseteq L_{\oR}^{\Le,\C}$, given by
\begin{equation}
  L_{X}^{\Le,\C}=\{\phi\in L_R^{\Le,\C} : \phi_{E,\ls,\ms}^{\Li}=0,\; \phi_{E,\ls,\ms}^{\Lci}=0\} .
  \label{eq:evandec}
\end{equation}
In contrast to the propagating case, $L_{X}^{\Le,\C}$ defines a real polarization rather than a Kähler polarization. In particular, $L_{X}^{\Le,\C}$ is a null rather than positive-definite subspace for the sesquilinear form (\ref{eq:stdip}). Consequently, standard Kähler quantization is not applicable. Rather, we shall use the novel $\alpha$-Kähler quantization scheme developed for this purpose in \cite{CoOe:locgenvac}.

Apart from a choice of vacuum in the exterior of the hypercylinder we also need a choice of vacuum in the interior. It turns out that it is convenient to let the vacuum be determined by the local behavior of solutions near the hypersurface. As we have seen above, in the vicinity of the hypersurface an exponential decay to the interior is exhibited by the modified spherical Bessel functions $\tilde{k}_{\ls}$, compare (\ref{eq:asymptk}). Therefore, we take the corresponding Lagrangian subspace to determine the interior vacuum. We denote this by
\begin{equation}
  L_{<R}^{\Le,\C}=\{\phi\in L_R^{\Le,\C} : \phi_{E,\ls,\ms}^{\Lx}=0,\; \phi_{E,\ls,\ms}^{\Lcx}=0\} .
  \label{eq:evaninc}
\end{equation}

The exterior and interior vacua, both for the propagating and evanescent sectors, can be encoded equivalently in terms of a complex structure $J_{\oR}:L_R^{\C}\to L_R^{\C}$. This is characterized by having eigenvalues $\im$ and $-\im$ on the corresponding Lagrangian subspaces, that is, on $L_{X}^{\C}$ and $\overline{L_{X}^{\Lp,\C}}\oplus L_{<R}^{\Le,\C}$ respectively. Explicitly, we have,
\begin{gather}
  (J_{\oR}\phi)_{E,\ls,\ms}^\Lout=\im \phi_{E,\ls,\ms}^\Lout, \;
  (J_{\oR}\phi)_{E,\ls,\ms}^\Lcin=\im \phi_{E,\ls,\ms}^\Lcin, \;
  (J_{\oR}\phi)_{E,\ls,\ms}^\Lcout=-\im \phi_{E,\ls,\ms}^\Lcout, \;
  (J_{\oR}\phi)_{E,\ls,\ms}^\Lin=-\im \phi_{E,\ls,\ms}^\Lin, \label{eq:cstrp} \\
  (J_{\oR}\phi)_{E,\ls,\ms}^\Lx=\im \phi_{E,\ls,\ms}^\Lx, \;
  (J_{\oR}\phi)_{E,\ls,\ms}^\Lcx=\im \phi_{E,\ls,\ms}^\Lcx, \;
  (J_{\oR}\phi)_{E,\ls,\ms}^\Li=-\im \phi_{E,\ls,\ms}^\Li, \;
  (J_{\oR}\phi)_{E,\ls,\ms}^\Lci=-\im \phi_{E,\ls,\ms}^\Lci .
\label{eq:cstre}
\end{gather}
Note that in the propagating sector, $J_{\oR}$ restricts to a complex structure on the real solutions space $L_R^{\Lp}\subseteq L_R^{\Lp,\C}$, while this is not the case in the evanescent sector.

The map $\alpha:L_R^{\Le,\C}\to L_R^{\Le,\C}$ required for the construction of the Hilbert space for the evanescent sector is a \emph{real structure}. That is, it behaves as a complex conjugation in the sense $\alpha(\im\phi)=-\im\alpha(\phi)$ and $\alpha^2(\phi)=\phi$. Moreover, $\alpha$ is required to be compatible with the symplectic form in the sense, $\omega_R(\alpha(\phi),\alpha(\xi))=\overline{\omega_R(\phi,\xi)}$. Also, $\alpha$ must interchange the polarized subspaces $L_{X}^{\Le,\C}$ and $L_{<R}^{\Le,\C}$ defining the vacua. Note that this implies that $\alpha$ commutes with $J_{\oR}$. Finally, $\alpha$ is required to make the sesquilinear form
\begin{equation}
  (\phi,\xi)_{\oR}^{\alpha}=4\im\omega_{\oR}(\alpha(\phi),\xi)
  \label{eq:alphaip}
\end{equation}
positive-definite on the polarized subspace $L_{X}^{\Le,\C}$ that defines the exterior vacuum.
Instead of looking directly for a positive-definite compatible real structure $\alpha$, it has turned out fruitful in the context of real polarizations to first consider a \emph{positive-definite reflection map} $\gamma:L_R^{\Le}\to L_R^{\Le}$ \cite{CoOe:locgenvac,CoOe:evanescent}. This is required to have the following properties: $\gamma$ is an involution, i.e., $\gamma^2(\phi)=\phi$, $\gamma$ is compatible with the symplectic form in the sense $\omega_R(\gamma(\phi),\gamma(\xi))=-\omega_R(\phi,\xi)$, and $\gamma$ interchanges the polarized subspaces $L_{X}^{\Le}$ and $L_{<R}^{\Le}$. Finally, $\gamma$ must satisfy the positivity condition, $\omega_{\oR}(\overline{\gamma(\phi)},\phi)>0$ if $\phi\in L_{X}^{\Le}\setminus\{0\}$.
Given a positive-definite reflection map $\gamma$, the map $\alpha(\phi)=-\im\overline{\gamma(\phi)}$ is then automatically a positive-definite compatible real structure.

In the present context we can make further demands on $\gamma$ which will then also be satisfied by $\alpha$. The first demand is the \emph{interior compatibility condition}. This means that $\gamma$ leaves invariant the subspace $L_{M}^{\Le,\C}\subseteq L_R^{\Le,\C}$ that corresponds to solutions regular in the interior of the solid hypercylinder. A consequence of this is that the calculation of amplitudes and correlation functions is quite analogous to the usual Kähler quantization case. The second demand we shall make is that $\gamma$ is covariant (or equivariant as a mathematician would say) with respect to spacetime isometries of the hypercylinder: rotations and time translations. It turns out that there is a single map $\gamma$ that satisfies all these conditions. This is,
\begin{gather}
  (\gamma(\xi))^{\Lx}_{E,\ls,\ms}=-\xi^{\Li}_{E,\ls,\ms},\;
  (\gamma(\xi))^{\Lcx}_{E,\ls,\ms}=-\xi^{\Lci}_{E,\ls,\ms},\;
  (\gamma(\xi))^{\Li}_{E,\ls,\ms}=-\xi^{\Lx}_{E,\ls,\ms},\;
  (\gamma(\xi))^{\Lci}_{E,\ls,\ms}=-\xi^{\Lcx}_{E,\ls,\ms}, \\
  (\alpha(\xi))_{E,\ls,\ms}^{\Lx}=\im\, \overline{\xi_{E,\ls,\ms}^{\Lci}},\;  (\alpha(\xi))_{E,\ls,\ms}^{\Lcx}=\im\, \overline{\xi_{E,\ls,\ms}^{\Li}},\;  (\alpha(\xi))_{E,\ls,\ms}^{\Li}=\im\, \overline{\xi_{E,\ls,\ms}^{\Lcx}},\;
  (\alpha(\xi))_{E,\ls,\ms}^{\Lci}=\im\, \overline{\xi_{E,\ls,\ms}^{\Lx}} .
\end{gather}
The inner product (\ref{eq:alphaip}) is then given by
\begin{equation}
  (\phi,\xi)_{\oR}^{\Le,\alpha}=
  \int_{0}^{m}\xd E\frac{p}{2\pi}\sum_{\ls,\ms}
  \left(\xi_{E,\ls,\ms}^\Lx\overline{\phi_{E,\ls,\ms}^\Lx}
  +\xi_{E,\ls,\ms}^\Lcx\overline{\phi_{E,\ls,\ms}^\Lcx}
  -\xi_{E,\ls,\ms}^\Li\overline{\phi_{E,\ls,\ms}^\Li}
  -\xi_{E,\ls,\ms}^\Lci\overline{\phi_{E,\ls,\ms}^\Lci}
  \right) .
\end{equation}
We also exhibit the bilinear form corresponding to (\ref{eq:pbilin}). Decomposing $L_R^{\Le,\C}=L_{X}^{\Le,\C} \oplus L_{<R}^{\Le,\C}$ with notation $\phi=\phi^+ + \phi^-$ we have
\begin{equation}
  \{\phi,\xi\}_{\oR}^{\Le}=4\im\omega_{\oR}^{\Le}(\phi^-,\xi^+)
  =-\int_0^{m}\xd E\frac{\im p}{2\pi}\sum_{\ls,\ms}
  \left(\xi_{E,\ls,\ms}^\Lx\phi_{E,\ls,\ms}^\Lci
  + \xi_{E,\ls,\ms}^\Lcx\phi_{E,\ls,\ms}^\Li\right) .
  \label{eq:ebilin}
\end{equation}
This is positive-definite on the $\alpha$-twisted real solution space $L_R^{\Le,\alpha}=\{\phi\in L_R^{\Le,\C} : \alpha(\phi)=\phi\}$.

Having determined the positive-definite compatible real structure $\alpha$, uniquely fixes the construction of the Hilbert space $\cH_R^{\Le}$ of states of the evanescent sector associated to the hypercylinder at radius $R$. The creation and annihilation operators are labeled by elements of $L_R^{\Le,\alpha}$, and satisfy the commutation relations analogous to the propagating sector (\ref{eq:pcom}),
\begin{equation}
  [a_\eta,a_\phi^\dagger]=\{\phi,\eta\}_{\oR}^{\Le} .
  \label{eq:ecom}
\end{equation}
The total Hilbert space of states associated to the hypercylinder is the (completed) tensor product of the Hilbert spaces for the propagating and evanescent sectors, $\cH_R=\cH_R^{\Lp}\tens\cH_R^{\Le}$.

For the evanescent sector, the parametrization of creation and annihilation operators as well as coherent states (see below) in terms of elements of the twisted phase space $L_R^{\Le,\alpha}$ instead of the ordinary phase space $L_R^{\Le}$ opens the question as to how a “correct” correspondence between classical and quantum degrees of freedom can be established. A possible step in resolving this consists in introducing a linear identification map between the usual and the twisted phase space $I^{\Le}:L_R^{\Le}\to L_R^{\Le,\alpha}$. In the present context, such an identification map, satisfying several desirable properties is given by a multiple of the projection map $P_R^{\Le,\alpha}:L_R^{\Le,\C}\to L_R^{\Le,\alpha}$, restricted to $L_R^{\Le}$ \cite{CoOe:locgenvac}:
\begin{equation}
  I_R^{\Le}(\phi)= \sqrt{2} P_R^{\Le,\alpha}(\phi)\qquad\forall\phi\in L_R^{\Le} .
  \label{eq:idintcompat}
\end{equation}
As shown in \cite{CoOe:locgenvac}, when $\alpha$ is derived from a positive-definite reflection map as here, the map $I_R^{\Le}$ defined in this way is indeed an isomorphism. What is more, $I_R^{\Le}$ inherits from $\alpha$ the interior compatibility condition in the sense,
\begin{equation}
  I_R^{\Le}(L_M^{\Le})=L_M^{\Le,\alpha} .
\end{equation}
$I_R^{\Le}$ extends uniquely to a complex linear isomorphism $L_R^{\Le,\C}\to L_R^{\Le,\C}$ as follows,
\begin{equation}
  I_R^{\Le}(\phi)
  =\frac{1}{\sqrt{2}}(\phi+\alpha(\overline{\phi}))
  =\frac{1}{\sqrt{2}}(\phi-\im\gamma(\phi))
   \qquad \forall \phi\in L_R^{\Le,\C} .
   \label{eq:idralpha}
\end{equation}
Explicitly, this is here,
\begin{gather}
  \left(I_R^{\Le}(\phi)\right)^{\Lx}_{E,\ls,\ms}
  =\frac{1}{\sqrt{2}}\left(\phi^{\Lx}_{E,\ls,\ms}
  +\im \phi^{\Li}_{E,\ls,\ms}\right), \quad
  \left(I_R^{\Le}(\phi)\right)^{\Li}_{E,\ls,\ms}
  =\frac{1}{\sqrt{2}}\left(\phi^{\Li}_{E,\ls,\ms}
  +\im \phi^{\Lx}_{E,\ls,\ms}\right), \nonumber \\
  \left(I_R^{\Le}(\phi)\right)^{\Lcx}_{E,\ls,\ms}
  =\frac{1}{\sqrt{2}}\left(\phi^{\Lcx}_{E,\ls,\ms}
  +\im \phi^{\Lci}_{E,\ls,\ms}\right), \quad
  \left(I_R^{\Le}(\phi)\right)^{\Lci}_{E,\ls,\ms}
  =\frac{1}{\sqrt{2}}\left(\phi^{\Lci}_{E,\ls,\ms}
  +\im \phi^{\Lcx}_{E,\ls,\ms}\right) .
\end{gather}
The inverse $I_R^{\Le\, -1}$ of $I_R^{\Le}$ is given by,
\begin{equation}
  I_R^{\Le\, -1}(\phi)
  =\frac{1}{\sqrt{2}}(\phi-\alpha(\overline{\phi}))
  =\frac{1}{\sqrt{2}}(\phi+\im\gamma(\phi))
   \qquad \forall \phi\in L_R^{\Le,\C} .
   \label{eq:idrialpha}
\end{equation}
We can now pull back the complex structure $J_{\oR}^{\Le}$ from $L_R^{\Le,\alpha}$ to a complex structure $\tilde{J}_{\oR}^{\Le}$ on $L_R^{\Le}$ via $\tilde{J}_{\oR}^{\Le}=I_R^{\Le\, -1} J_{\oR}^{\Le} I_R^{\Le}$. Explicitly, this is,
\begin{equation}
  (\tilde{J}_{\oR}^{\Le}\phi)_{E,\ls,\ms}^{\Lx}=-\phi_{E,\ls,\ms}^{\Li},\;
  (\tilde{J}_{\oR}^{\Le}\phi)_{E,\ls,\ms}^{\Lcx}=-\phi_{E,\ls,\ms}^{\Lci},\;
  (\tilde{J}_{\oR}^{\Le}\phi)_{E,\ls,\ms}^{\Li}=\phi_{E,\ls,\ms}^{\Lx},\;
  (\tilde{J}_{\oR}^{\Le}\phi)_{E,\ls,\ms}^{\Lci}=\phi_{E,\ls,\ms}^{\Lcx} .
\end{equation}

By construction, the complex structure we obtain restricts to the real solution space $L_R^{\Le}\subseteq L_R^{\Le,\C}$. This signals that we can use it to perform an ordinary Kähler quantization of $L_R^{\Le}$ to obtain a Hilbert space $\tilde{\cH}_R^{\Le}$. This is precisely what was done in \cite[Section~5.3]{Oe:holomorphic} with the exact same complex structure $\tilde{J}_{\oR}^{\Le}$ that we have just obtained (together with the conventional quantization of $L_R^{\Lp}$ to $\cH_R^{\Lp}$ as reviewed above). From the present perspective we understand that this quantization is “wrong” in the sense of not being based on the physical vacuum. In particular, this will not lead to the correct correlation functions. However, we also understand now \cite{CoOe:locgenvac} that the two quantizations are related in a simple way, precisely through the identification map $I_R^{\Le}$. In particular, there is a unitary map $\tilde{U}_R^{\Le}:\tilde{\cH}_R^{\Le}\to\cH_R^{\Le}$, to be made explicit in Section~\ref{sec:cohsobs}.


\section{Particle states}
\label{sec:particles}

In this section we consider particle states on the hypercylinder. In global QFT in Minkowski space, often the most convenient parametrization of particle states is in terms of 3-momenta. In contrast, a natural parametrization in the present setting is given in terms of energy and angular momentum quantum numbers. Moreover, an additional binary quantum number arises. In the propagating case this corresponds to the particle being incoming or outgoing with respect to the hypercylinder \cite{Oe:kgtl}, compare (\ref{eq:inmodes})  and (\ref{eq:outmodes}).
In the evanescent case this can be identified with the choice between exponential decay to the exterior or to the interior of the hypercylinder.

We consider propagating particles first. We introduce field modes $\Phi^{\Lin,E,\ls,\ms},\Phi^{\Lout,E,\ls,\ms}\in L_R^{\Lp}$ to model particles with the corresponding quantum numbers. In terms of the field parametrization (\ref{eq:pparam}), these are given by,
\begin{align}
    (\Phi^{\Lout,E,\ls,\ms})_{E',\ls',\ms'}^\Lout
    = (\Phi^{\Lout,E,\ls,\ms})_{E',\ls',\ms'}^\Lcout
    & =\sqrt{\frac{2\pi}{p}}\delta_{\ls,\ls'}\delta_{\ms,\ms'}\delta(E-E') ,
    \label{eq:partpin}  \\
    (\Phi^{\Lin,E,\ls,\ms})_{E',\ls',\ms'}^\Lin
    = (\Phi^{\Lin,E,\ls,\ms})_{E',\ls',\ms'}^\Lcin
    & =\sqrt{\frac{2\pi}{p}}\delta_{\ls,\ls'}\delta_{\ms,\ms'}\delta(E-E') .
    \label{eq:partpout}
\end{align}
The other coefficients are zero. With (\ref{eq:pbilin})  and (\ref{eq:pcom}) we find that the corresponding creation and annihilation operators satisfy the commutation relations,
\begin{align}
    [a_{\Lin,E,\ls,\ms},a^\dagger_{\Lin,E',\ls',\ms'}]=\delta_{\ls,\ls'}\delta_{\ms,\ms'} \delta(E-E'), \label{eq:ccrin} \\
    [a_{\Lout,E,\ls,\ms},a^\dagger_{\Lout,E',\ls',\ms'}]=\delta_{\ls,\ls'}\delta_{\ms,\ms'} \delta(E-E') . \label{eq:ccrout}
\end{align}
Commutators involving both incoming and outgoing particles vanish.

We proceed to consider the evanescent particles. For these, we indicate the quantum number corresponding with exponential decay either to the outside or to the inside by (x) and (i) respectively.
We introduce field modes $\Phi^{\Lx,E,\ls,\ms},\Phi^{\Li,E,\ls,\ms}\in L_R^{\Le}$ to model particles with the corresponding quantum numbers. In terms of the field parametrization (\ref{eq:eparam}), we take these to be given by,
\begin{align}
    (\Phi^{\Lx,E,\ls,\ms})_{E',\ls',\ms'}^{\Lx} & = (\Phi^{\Lx,E,\ls,\ms})_{E',\ls',\ms'}^{\Lcx} =\sqrt{\frac{2\pi}{p}}\delta_{\ls,\ls'}\delta_{\ms,\ms'}\delta(E-E') \\
    (\Phi^{\Li,E,\ls,\ms})_{E',\ls',\ms'}^{\Li} & = \im\sqrt{\frac{2\pi}{p}}\delta_{\ls,\ls'}\delta_{\ms,\ms'}\delta(E-E'), \; (\Phi^{\Li,E,\ls,\ms})_{E',\ls',\ms'}^{\Lci} =-\im\sqrt{\frac{2\pi}{p}}\delta_{\ls,\ls'}\delta_{\ms,\ms'}\delta(E-E') .
\end{align}
The other coefficients are zero.
These modes can not be directly considered for quantization as they live in the real phase space $L_{R}^{\Le}$ rather than in the $\alpha$-twisted phase space $L_{R}^{\Le,\alpha}$. We use the identification map $I^{\Le}$ given by (\ref{eq:idralpha}) to obtain the corresponding elements in $L_{R}^{\Le,\alpha}$ and denote them with a tilde,
\begin{align}
    (\tilde{\Phi}^{\Lx,E,\ls,\ms})_{E',\ls',\ms'}^{\Lx} & = (\tilde{\Phi}^{\Lx,E,\ls,\ms})_{E',\ls',\ms'}^{\Lcx} ={\sqrt\frac{\pi}{p}}\delta_{\ls,\ls'}\delta_{\ms,\ms'}\delta(E-E') , \nonumber \\
    (\tilde{\Phi}^{\Lx,E,\ls,\ms})_{E',\ls',\ms'}^{\Li} & = (\tilde{\Phi}^{\Lx,E,\ls,\ms})_{E',\ls',\ms'}^{\Lci} =\im\sqrt{\frac{\pi}{p}}\delta_{\ls,\ls'}\delta_{\ms,\ms'}\delta(E-E') , \label{eq:partedec}\\
    (\tilde{\Phi}^{\Li,E,\ls,\ms})_{E',\ls',\ms'}^{\Li} & = \im\sqrt{\frac{\pi}{p}}\delta_{\ls,\ls'}\delta_{\ms,\ms'}\delta(E-E'), \; (\tilde{\Phi}^{\Li,E,\ls,\ms})_{E',\ls',\ms'}^{\Lci} =-\im\sqrt{\frac{\pi}{p}}\delta_{\ls,\ls'}\delta_{\ms,\ms'}\delta(E-E') , \nonumber \\
    (\tilde{\Phi}^{\Li,E,\ls,\ms})_{E',\ls',\ms'}^{\Lx} & = -\sqrt{\frac{\pi}{p}}\delta_{\ls,\ls'}\delta_{\ms,\ms'}\delta(E-E'), \; (\tilde{\Phi}^{\Li,E,\ls,\ms})_{E',\ls',\ms'}^{\Lcx} =\sqrt{\frac{\pi}{p}}\delta_{\ls,\ls'}\delta_{\ms,\ms'}\delta(E-E') . \label{eq:parteinc}
\end{align}
Using the obvious notation for the corresponding creation and annihilation operators, from (\ref{eq:ebilin}) and (\ref{eq:ecom}), these satisfy the commutation relations,
\begin{align}
    [a_{\Lx,E,\ls,\ms},a^\dagger_{\Lx,E',\ls',\ms'}]=\delta_{\ls,\ls'}\delta_{\ms,\ms'} \delta(E-E'), \label{eq:ccrx} \\
    [a_{\Li,E,\ls,\ms},a^\dagger_{\Li,E',\ls',\ms'}]=\delta_{\ls,\ls'}\delta_{\ms,\ms'} \delta(E-E') . \label{eq:ccri}
\end{align}

The commutation relations (\ref{eq:ccrin}), (\ref{eq:ccrout}), (\ref{eq:ccrx}), and (\ref{eq:ccri}), lead to a simple \emph{completeness relation} for the 1-particle subspace $\cH_R^1\subseteq \cH_R$,
\begin{equation}
    \id^1=\sum_{l,m}\left(\int_{m}^{\infty}\xd E\,
      \left(P_{\Lin,E,\ls,\ms} + P_{\Lout,E,\ls,\ms}\right) + \int_{0}^m \xd E\,
      \left(P_{\Lx,E,\ls,\ms} + P_{\Li,E,\ls,\ms}\right)\right)
\end{equation}
Here, $P_{\bullet,E,\ls,\ms}$ represents the projector onto the corresponding state,\footnote{Strictly speaking, the operators $P_{\bullet,E,\ls,\ms}$ are not projection operators due to the singular nature of the commutation relations with respect to the energy variable.}
\begin{equation}
   P_{\bullet,E,\ls,\ms} =a^\dagger_{\bullet,E,\ls,\ms}|0\rangle
    \langle 0 | a_{\bullet,E,\ls,\ms} .
\end{equation}
The completeness relation extends straightforwardly to the $n$-particle sector of the state space. In that case there will be $n$ sums and integrals over the energy and angular momentum quantum numbers.


\section{Coherent states and the action of observables}
\label{sec:cohsobs}

In order to describe the quantization of observables, it is convenient to use \emph{coherent states}. The latter are obtained by acting with exponentiated creation operators on the vacuum state, and they generate a dense subspace of the Hilbert space. In usual Kähler quantization, coherent states are labeled by elements of the phase space and behave in many ways as approximations of the respective classical phase space element. This is true also on the hypercylinder in the propagating sector. In the evanescent sector, the coherent states are labeled instead by elements of the $\alpha$-twisted phase space. Here, a good understanding of the classical-quantum correspondence is still lacking. Thus, a coherent state $\coh_\xi\in\cH_R$ is labeled by an element of $\xi\in L_R^{\Lp}\oplus L_R^{\Le,\alpha}$. It may be obtained as,
\begin{equation}
  \coh_\xi=\exp\left(\frac{1}{\sqrt{2}} a_\xi^\dagger\right)\coh_0.
\end{equation}
Here, $\coh_0$ is the vacuum state and at the same time the coherent state associated to the null vector.
The inner product between coherent states is,
\begin{equation}
  \langle \coh_\xi,\coh_\phi\rangle_R =\exp\left(\frac12\{\phi,\xi\}_{\overline{R}}\right) .
  \label{eq:cohip}
\end{equation}
In the following we will also consider \emph{normalized} coherent states, given by,
\begin{equation}
  \ncoh_\xi=\exp\left(-\frac14 \{\xi,\xi\}_{\overline{R}}\right) \coh_\xi .
\end{equation}
In the evanescent sector, coherent states in the Hilbert space $\tilde{\cH}_R$ with the alternative quantization are labeled by elements of the real phase space $L_R^{\Le}\subseteq L_R^{\Le,\C}$. The unitary map $\tilde{U}_R^{\Le}:\tilde{\cH}_R^{\Le}\to\cH_R^{\Le}$ takes on coherent states the simple form,
\begin{equation}
  \tilde{U}_R^{\Le}(\tilde{\coh}_\xi)=\coh_{I_R^{\Le}(\xi)} .
\end{equation}

Classical observables are functions on the phase space, which in this case is the space $L_{R}$ of germs of solutions on the hypercylinder. To have a mathematically well-behaved notion of observable we shall require these to extend to holomorphic functions on the complexified phase space $L_{R}^{\C}$. An observable $F$ is then real if it satisfies the condition
\begin{equation}
  F(\overline{\phi})=\overline{F(\phi)} .
  \label{eq:realobs}
\end{equation}
In Kähler quantization, real observables give rise to Hermitian operators on the Hilbert space of states. In $\alpha$-Kähler quantization this is no longer true. Instead, $\alpha$-real observables give rise to Hermitian operators. They satisfy the $\alpha$-twisted realness condition,
\begin{equation}
  F(\alpha(\phi))=\overline{F(\phi)} .
  \label{eq:alphaobs}
\end{equation}
In the present case, for the propagating sector $L_{R}^{\Lp,\C}$ we have the first condition (\ref{eq:realobs}), while for the evanescent sector $L_{R}^{\Le,\C}$ we have the second condition (\ref{eq:alphaobs}). We can also subsume both conditions into one by setting $\alpha(\phi)=\overline{\phi}$ for propagating solutions $\phi\in L_{R}^{\Lp,\C}$.

In order to discuss the action of observables on the Hilbert space $\cH_{R}$ it will be convenient to restrict to \emph{Weyl observables}, which are generators for a large class of observables, including polynomial ones. Weyl observables arise a follows. Associated to $\xi\in L_{R}^\C$ we define the linear observable $D_\xi:L_{R}^\C\to\C$ and the Weyl observable $F_\xi:L_{R}^\C\to\C$ by,
\begin{equation}
  D_\xi(\phi)\defeq 2\omega_{\overline{R}}(\xi,\phi), \qquad
  F_\xi\defeq \exp(\im D) .
\end{equation}
The action of the quantized Weyl observable $\hat{F}_\xi$ on a coherent state $\coh_\phi$ is \cite{CoOe:locgenvac},
\begin{equation}
  \hat{F}_\xi \coh_\phi=\exp\left(-\frac12 \{\phi,\xi\}_{\overline{R}}-\frac14 \{\xi,\xi\}_{\overline{R}}\right) \coh_{\phi+\xi^- +\alpha(\xi^-)} .
\end{equation}
As above, we have used here the decomposition $\xi=\xi^{+} +\xi^{-}$ corresponding to $L_{R}^{\C}=L_{R}^{+}\oplus L_{R}^{-}$ with $L_{R}^{+}=L_{X}^{\Lp,\C}\oplus L_{X}^{\Le,\C}$ and $L_{R}^{-}=\overline{L_{X}^{\Lp,\C}}\oplus L_{<R}^{\Le,\C}$. The quantized Weyl observables satisfy the \emph{Weyl relations},
\begin{equation}
  \hat{F_\xi} \hat{F_\phi}=\exp(\im\omega_{\overline{R}}(\xi,\phi))\hat{F}_{\xi+\phi} .
\end{equation}
If $\xi$ is $\alpha$-real, i.e., $\xi\in L_{R}^{\Lp} \oplus L_{R}^{\Le,\alpha}$ so that $D_\xi$ is also $\alpha$-real, the action of $\hat{F}_\xi$ becomes unitary. On normalized coherent states we then have,
\begin{equation}
  \hat{F_\xi} \ncoh_\phi=\exp\left(\im\omega_{\overline{R}}(\xi,\phi)\right) \ncoh_{\phi+\xi} .
\end{equation}
In particular, this implies, as previously stated, that an $\alpha$-real observable generated from a Weyl observable acts as a Hermitian operator.


\section{Radial evolution}
\label{sec:radevol}

In this section we consider the radial evolution of states between hypercylinders of different radii. More specifically, we shall consider the evolution from a larger radius $R$ to a smaller radius $R'$. We shall denote the corresponding evolution map between the Hilbert spaces of states by $V_{[R,R']}:\cH_R\to\cH_{R'}$. The evolution map is related to the amplitude map $\rho_{[R,R']}:\cH_{R}\tens\cH_{\overline{R'}}\to\C$ for the region between the two hypercylinders as follows,
\begin{equation}
  \langle \psi', V_{[R,R']} \psi\rangle_{R'}
  =\rho_{[R,R']}(\psi\tens\iota(\psi')) ,
\end{equation}
where $\psi\in\cH_R$ and $\psi'\in\cH_{R'}$. Evaluated on coherent states, the amplitude is (\cite{CoOe:locgenvac}, analogous to \cite{CoOe:evanescent}),
\begin{equation}
    \langle \coh_{\xi'}, V_{[R,R']} \coh_{\xi}\rangle_{R'}
    =\rho_{[R,R']}(\coh_{\xi}\tens\coh_{\xi'})
    =\exp\left(\frac12 \{\xi,\xi'\}_{\oR}\right) .
    \label{eq:revolampl}
\end{equation}
We have taken advantage here of the fact that the spaces $L_R^{\C}$ and $L_{R'}^{\C}$ are canonically isomorphic, and the same is true for the subspaces $L_R^{\alpha}$ and $L_{R'}^{\alpha}$. We may thus recognize the right-hand side of (\ref{eq:revolampl}) as the inner product of coherent states, compare equation (\ref{eq:cohip}). This means that the radial evolution map $V_{[R,R']}$ simply is the identity under the identification of corresponding coherent states, i.e.,
\begin{equation}
    V_{[R,R']}\coh_\xi=\coh_\xi .
\end{equation}
In particular, radial evolution in the present quantization is unitary, not only on the propagating sector, but also on the evanescent sector. This is in contrast to the corresponding analysis of the situation for evolution between hyperplanes \cite{CoOe:evanescent}.


\section{Scattering amplitudes of the free field}
\label{sec:ampl}

Scattering events in quantum field theory are usually described through the S-matrix, which is an asymptotic version of a transition amplitude. The transition amplitude encodes information about the processes taking place between the initial and the final time (in all of space), given an initial and a final state. While this is the conventional approach, there is an alternative approach made possible in the present framework of GBQFT. In this approach we consider the amplitude for a \emph{single} state on the hypercylinder \cite{Oe:kgtl}. This amplitude encodes information about the processes in the interior of the hypercylinder, that is, inside a sphere of radius $R$ in space, at all times. We may then take a limit of letting the radius go to infinity and obtain an asymptotic amplitude that is in fact equivalent to the usual S-matrix \cite{CoOe:spsmatrix,CoOe:smatrixgbf}.

A disadvantage of the asymptotic description of scattering is that we miss finite-size effects such as fields that decay too quickly as distance increases. This is precisely the case of evanescent waves. While this is difficult to remedy in the traditional approach, the hypercylinder approach is ideal to address this situation. We merely need to consider the amplitude at finite radius $R$, without taking the limit. Previously, this amplitude was only partially known, sufficient for the asymptotic limit, but excluding the evanescent sector. We present in the following the complete amplitude as well as corresponding correlation functions.

Using the complex structure (\ref{eq:cstrp}) and (\ref{eq:cstre}) we define the bilinear form
\begin{multline}
  g_{\oR}(\phi,\xi)\defeq 2\omega_{\oR}(\phi,J_{\oR}\xi)= \\
  \int_{m}^\infty\xd E\frac{p}{4\pi}\sum_{\ls,\ms}
  \left(\xi_{E,\ls,\ms}^\Lout \phi_{E,\ls,\ms}^\Lcout
  +\xi_{E,\ls,\ms}^\Lcin \phi_{E,\ls,\ms}^\Lin
  +\xi_{E,\ls,\ms}^\Lcout \phi_{E,\ls,\ms}^\Lout
  +\xi_{E,\ls,\ms}^\Lin \phi_{E,\ls,\ms}^\Lcin
  \right) \\
  - \int_{0}^m\xd E\frac{\im p}{4\pi}\sum_{\ls,\ms}
  \left(\xi_{E,\ls,\ms}^\Lx\phi_{E,\ls,\ms}^\Lci+\xi_{E,\ls,\ms}^\Lcx\phi_{E,\ls,\ms}^\Li
  +\xi_{E,\ls,\ms}^\Li\phi_{E,\ls,\ms}^\Lcx+\xi_{E,\ls,\ms}^\Lci\phi_{E,\ls,\ms}^\Lx\right) .
  \label{eq:ipg}
\end{multline}
This bilinear form is also the same as the symmetric part of the bilinear forms (\ref{eq:pbilin}) and (\ref{eq:ebilin}). Note in particular that $g_{\oR}$ is a real positive-definite inner product on $L_R^{\alpha}$. Crucially, the complex structure induces a direct sum decomposition $L_R^{\alpha}=L_M^{\alpha}\oplus J_{\oR} L_M^{\alpha}$, orthogonal with respect to $g_{\oR}$ \cite{Oe:holomorphic}. We use the notation $\phi=\phi^{\text{R}}+J_{\oR}\phi^{\text{I}}$ for $\phi\in L_{R}^{\alpha}$ with $\phi^{\text{R}},\phi^{\text{I}}\in L_M^{\alpha}$. This is a decomposition into solutions that are regular in the interior of the hypercylinder and a certain complement of solutions that are not.
To be more explicit, for the propagating sector, while the subspace $L_M^{\Lp,\C}$ is given by (\ref{eq:propint}) its complement $J_{\overline{R}} L_M^{\Lp,\C}$ is given by,
\begin{equation}
  J_{\overline{R}} L_M^{\Lp,\C}
  =\{\phi\in L_R^{\Lp,\C}:
   \phi_{E,\ls,\ms}^\Lin=-\phi_{E,\ls,\ms}^\Lout,
   \phi_{E,\ls,\ms}^\Lcin=-\phi_{E,\ls,\ms}^\Lcout \} .
   \label{eq:propcomp}
\end{equation}
The components $\xi^{\text{R}}$ and $J_{\oR}\xi^{\text{I}}$ consist thus of solutions that are linear combinations of solutions of the form (as well as their complex conjugates),
\begin{equation}
  j_{\ls}(pr) e^{-\im E t}Y_{\ls}^{\ms}(\Omega)\qquad\text{and respectively}\quad
  n_{\ls}(pr) e^{-\im E t}Y_{\ls}^{\ms}(\Omega) .
\end{equation}
For the evanescent sector the subspace $L_M^{\Le,\C}$ is given by (\ref{eq:evanint}) while its complement $J_{\overline{R}} L_M^{\Le,\C}$ is given by,
\begin{equation}
  J_{\overline{R}} L_{M}^{\Le,\C}=\{\phi\in L_R^{\Le,\C} : \phi_{E,\ls,\ms}^{\Li}=(-1)^{\ls+1}\phi_{E,\ls,\ms}^{\Lx},\; \phi_{E,\ls,\ms}^{\Lci}=(-1)^{\ls+1}\phi_{E,\ls,\ms}^{\Lcx}\} .
  \label{eq:evancomp}
\end{equation}
The components $\xi^{\text{R}}$ and $J_{\oR}\xi^{\text{I}}$ consist of solutions that are linear combinations of solutions of the form,
\begin{equation}
  a_{\ls}(pr) e^{-\im E t}Y_{\ls}^{\ms}(\Omega)\qquad\text{and respectively}\quad
  \tilde{a}_{\ls}(pr) e^{-\im E t}Y_{\ls}^{\ms}(\Omega) ,
\end{equation}
as well as their complex conjugates.

The amplitude for normalized coherent states $\ncoh_\xi\in\cH_{R}$ is given by the formula \cite{Oe:holomorphic,CoOe:locgenvac}
\begin{equation}
  \rho_M(\ncoh_\xi)=\exp\left(-\frac{\im}{2}g_{\oR}(\xi^{\text{R}},\xi^{\text{I}})-\frac12 g_{\oR}(\xi^{\text{I}},\xi^{\text{I}})\right) .
  \label{eq:amplhcyl}
\end{equation}
In the case of standard Kähler quantization this formula has a remarkably simple and compelling interpretation \cite{Oe:holomorphic}. If the classical solution $\xi$ in a neighborhood of the boundary of the hypercylinder admits a continuation to the interior, then $\xi=\xi^{\text{R}}$ and $\xi^\text{I}=0$ and the argument of the exponential vanishes, yielding unit amplitude. If we switch on a classically forbidden (i.e., non-continuable) component $\xi^{\text{I}}$ we obtain an exponential suppression of the amplitude due to the second term in the argument, which becomes negative. Also, we obtain a phase due to the first term in the argument, which becomes imaginary. This is precisely the kind of behavior that we expect from quantum tunneling phenomena which is how we can interpret the present setting.
In the present case this interpretation applies fully to the propagating sector of the phase space. For the evanescent sector we can observe an analogous behavior, however, this is now not with respect to real solutions, but complexified ones. This complicates the interpretation as the simple correspondence between coherent states and real classical solutions no longer holds.

Taking advantage of the identification map (\ref{eq:idralpha}), we can apparently restore the desired correspondence for the evanescent sector \cite{CoOe:locgenvac}. In order to obtain formulas valid in both sectors, we set $I_R(\phi)=\phi^{\Lp}+I_R^{\Le}(\phi^{\Le})$ and $\tilde{J}_{\oR}(\phi)=J_{\oR}^{\Lp}(\phi^{\Lp})+\tilde{J}_{\oR}^{\Le}(\phi^{\Le})$. That is, no modification is applied to the propagating sector. The modified complex structure introduces a modified inner product $\tilde{g}_{\oR}$, analogous to (\ref{eq:ipg}), but now positive-definite on the real phase space $L_R$. We only exhibit its evanescent part, as the propagating one is unchanged from (\ref{eq:ipg}), by construction,
\begin{multline}
  \tilde{g}_{\oR}^{\Le}(\phi,\xi)\defeq 2\omega_{\oR}^{\Le}(\phi,\tilde{J}_{\oR}\xi) \\
  = \int_0^{m}\xd E\frac{p}{4\pi}\sum_{\ls,\ms}
  \left(\xi_{E,\ls,\ms}^\Li\phi_{E,\ls,\ms}^\Lci+\xi_{E,\ls,\ms}^\Lci\phi_{E,\ls,\ms}^\Li
  +\xi_{E,\ls,\ms}^\Lx\phi_{E,\ls,\ms}^\Lcx+\xi_{E,\ls,\ms}^\Lcx\phi_{E,\ls,\ms}^\Lx\right) .
\end{multline}
The complex structure $\tilde{J}_{\oR}$ induces a direct sum decomposition $L_R=L_M\oplus \tilde{J}_{\oR} L_M$, orthogonal with respect to $\tilde{g}_{\oR}$. We use the notation $\phi=\phi^{\tilde{\text{R}}}+\tilde{J}_{\oR}\phi^{\tilde{\text{I}}}$ for $\phi\in L_{R}$ with $\phi^{\tilde{\text{R}}},\phi^{\tilde{\text{I}}}\in L_M$. We then obtain the amplitude formula,
\begin{equation}
  \rho_M(\ncoh_{I_R(\xi)})=\exp\left(-\frac{\im}{2}\tilde{g}_{\oR}(\xi^{\tilde{\text{R}}},\xi^{\tilde{\text{I}}})-\frac12 \tilde{g}_{\oR}(\xi^{\tilde{\text{I}}},\xi^{\tilde{\text{I}}})\right) .
  \label{eq:amplmod}
\end{equation}
If we think of the coherent state $\ncoh_{I_R(\xi)}$ as in correspondence to the classical solution $\xi$ we obtain in the evanescent sector the same compelling physical interpretation of the amplitude as in the propagating sector. We shall see in the following, however, that this correspondence has its limitations when observables are involved.

We recall that for the propagating sector, the amplitude map (\ref{eq:amplhcyl}) can be converted to a form that is more in line with the traditional approach to scattering theory in quantum field theory, where the S-matrix is thought of as a map from a Hilbert space of incoming particles to a Hilbert space of outgoing particles. Indeed, the decomposition $L_R^{\Lp}=L_R^{\Lin}\oplus L_R^{\Lout}$ induces a decomposition of the Hilbert space $\cH_R^{\Lp}=\cH_R^{\Lin}\tens \cH_R^{\Lout}$. This in turn allows to convert the amplitude map on the propagating sector $\rho_M^{\Lp}$ into a unitary map $V_R^{\Lp}:\cH_R^{\Lin}\to \cH_{\oR}^{\Lout}$ via,
\begin{equation}
  \langle \psi', V_R^{\Lp}\psi\rangle
  = \rho_R^{\Lp}(\psi\tens\iota(\psi')) .
  \label{eq:inoutdef}
\end{equation}
We refer the interested reader to \cite{Oe:kgtl}, where this was first explained, as well as to \cite{CoOe:spsmatrix,CoOe:smatrixgbf}, where this was used to show equivalence with the standard S-matrix in the asymptotic limit $R\to\infty$.
To obtain a more explicit expression for the scattering map $V_R^{\Lp}$, it is convenient to introduce the linear map $u^{\Lp}:L_R^{\Lp,\C}\to L_R^{\Lp,\C}$ characterized by the following property \cite{Oe:freefermi}: $u^{\Lp}$ is the identity on $L_M^{\Lp,\C}$ given by (\ref{eq:propint}) and minus the identity on $J_{\overline{R}} L_M^{\Lp,\C}$ given by (\ref{eq:propcomp}). Explicitly, we find,
\begin{equation}
  (u^{\Lp}(\xi))_{E,\ls,\ms}^\Lout=\xi_{E,\ls,\ms}^\Lin,\;
  (u^{\Lp}(\xi))_{E,\ls,\ms}^\Lcout=\xi_{E,\ls,\ms}^\Lcin,\;
  (u^{\Lp}(\xi))_{E,\ls,\ms}^\Lin=\xi_{E,\ls,\ms}^\Lout,\;
  (u^{\Lp}(\xi))_{E,\ls,\ms}^\Lcin=\xi_{E,\ls,\ms}^\Lcout .
\end{equation}
In particular, $u^{\Lp}$ interchanges the components in the decomposition $L_R^{\Lp,\C}=L_R^{\Lin,\C}\oplus L_R^{\Lout,\C}$. It then follows that the map $V_R^{\Lp}$ in expression (\ref{eq:inoutdef}) takes the simple form,
\begin{equation}
  V_R^{\Lp}(\coh_{\xi})=\coh_{u^{\Lp}(\xi)} .
\end{equation}
See Lemma~8.3 and Proposition~8.4 in \cite{Oe:freefermi}.

In the present setting at finite $R$, we also have the evanescent sector $L_R^{\Le}$. For evanescent waves there is no sense in which they are incoming or outgoing with respect to the interior of the hypercylinder. However, we may look for a decomposition of the phase space with similar properties. It turns out that the decomposition into exponentially increasing vs.\ decaying solutions in the sense $L^{\Le,\C}=L_{X}^{\Le,\C}\oplus L_{<R}^{\Le,\C}$ is suitable, compare (\ref{eq:evandec}) and (\ref{eq:evaninc}). We denote the corresponding decomposition of the Hilbert space of states by, $\cH_R^{\Le}=\cH_{R}^X\oplus \cH_R^{<R}$. 
First, we work out the map $u^{\Le}: L_R^{\Le,\C}\to L_R^{\Le,\C}$, which is the identity on $L_M^{\Le,\C}$ given by (\ref{eq:evanint}) and minus the identity on $J_{\overline{R}} L_M^{\Le,\C}$ given by (\ref{eq:evancomp}),
\begin{align}
  (u^{\Le}(\xi))_{E,\ls,\ms}^\Li & =(-1)^{\ls}\xi_{E,\ls,\ms}^\Lx,\;
  (u^{\Le}(\xi))_{E,\ls,\ms}^\Lci =(-1)^{\ls}\xi_{E,\ls,\ms}^\Lcx, \nonumber \\
  (u^{\Le}(\xi))_{E,\ls,\ms}^\Lx & =(-1)^{\ls}\xi_{E,\ls,\ms}^\Li,\;
  (u^{\Le}(\xi))_{E,\ls,\ms}^\Lcx =(-1)^{\ls}\xi_{E,\ls,\ms}^\Lci .
\end{align}
We may now note that $u^{\Le}$ indeed exchanges the components in the decomposition $L^{\Le,\C}=L_{X}^{\Le,\C}\oplus L_{<R}^{\Le,\C}$. Defining the map $V_R^{\Le}:\cH_R^{X}\to \cH_{\oR}^{<R}$ analogous to (\ref{eq:inoutdef}), yields, similarly to the propagating sector,
\begin{equation}
  V_R^{\Le}(\coh_{\xi})=\coh_{u^{\Le}(\xi)} .
\end{equation}
Here $\xi\in L_X^{\Le,\alpha}$. We proceed to switch to the setting were coherent states for the evanescent sector are labeled by real solutions. This is facilitated by the following facts. As a consequence of interior compatibility of $\alpha$, it commutes with $u^{\Le}$. This in turn implies that $J_{\overline{R}} L_M^{\Le,\C}=\tilde{J}_{\overline{R}} L_M^{\Le,\C}$ are the same space, even though the complex structures are different maps. Moreover, $u^{\Le}$ commutes with the isomorphism $I_R^{\Le}$ given by (\ref{eq:idralpha}), and with its inverse. We may now consider the map $\tilde{V}_R^{\Le}:\tilde{\cH}_R^{X}\to \tilde{\cH}_{\oR}^{<R}$ defined analogously to $V_R^{\Le}$. This is,
\begin{equation}
  \tilde{V}_R^{\Le}(\tilde{\coh}_{\xi})=\tilde{\coh}_{u^{\Le}(\xi)} .
\end{equation}
Again, we obtain precisely a quantum map corresponding to the decomposition of the phase space $L^{\Le,\C}=L_{X}^{\Le,\C}\oplus L_{<R}^{\Le,\C}$. However, note that the map $\tilde{V}_R^{\Le}$ is \emph{not} induced from the map $V_R^{\Le}$ by the unitary transformation $\tilde{U}_R^{\Le}$. This is because the decomposition $L^{\Le,\C}=L_{X}^{\Le,\C}\oplus L_{<R}^{\Le,\C}$ is not invariant under the isomorphism $I_R^{\Le}$.

Finally, we provide an explicit form of the amplitude formula (\ref{eq:amplmod}). We obtain for $\xi\in L_R$,
\begin{multline}
  \rho_M(\ncoh_{I_R(\xi)})=\exp\left(\int_{m}^\infty \xd E\, \frac{p}{8\pi}\sum_{\ls,\ms}\left(2\, \xi_{E,\ls,\ms}^\Lin\,\xi_{E,\ls,\ms}^\Lcout
   - \xi_{E,\ls,\ms}^\Lin\,\xi_{E,\ls,\ms}^\Lcin
   - \xi_{E,\ls,\ms}^\Lout\,\xi_{E,\ls,\ms}^\Lcout \right)\right . \\
  \left. + \int_0^{m} \xd E\, \frac{p}{8\pi}\sum_{\ls,\ms}\left(
    ((-1)^{\ls+1}\im -1)\xi_{E,\ls,\ms}^\Li\xi_{E,\ls,\ms}^\Lci
    +((-1)^{\ls}\im -1)\xi_{E,\ls,\ms}^\Lx\xi_{E,\ls,\ms}^\Lcx \right.\right.\\
    \left.\left.
    +(-1)^{\ls}\xi_{E,\ls,\ms}^\Li\xi_{E,\ls,\ms}^\Lcx
    +(-1)^{\ls}\xi_{E,\ls,\ms}^\Lx\xi_{E,\ls,\ms}^\Lci
    \right)\right) .
\end{multline}
As before, we set $I_R(\xi)=\xi$ for $\xi\in L_R^\Lp$.


\section{Correlation functions}
\label{sec:correlation}

We proceed to consider the insertion of observables, leading from amplitudes to \emph{correlation functions}. On the one hand these can be used to model certain types of measurements, on the other hand these can be used to introduce interactions. In this context, the observables are defined on field configurations, as is usual for path integral quantization, i.e., they are off-shell observables. More specifically, we restrict our attention again to \emph{Weyl observables} which can be used as generators for a large class of observables, including polynomial ones.

Denote the space of \emph{field configurations} inside the solid hypercylinder $M$ by $K_M$. Again, we require observables to extend to holomorphic functions on the complexification $K_M^{\C}$. Let $D:K_M^{\C}\to\C$ be a linear, but not necessarily real observable. We define $F\defeq \exp(\im D)$ as the corresponding Weyl observable. As before, we decompose $\xi\in L_R^{\C}$ as $\xi^{\text{R}}+ J_{\oR} \xi^{\text{I}}$. Note that $\hat{\xi}\defeq \xi^{\text{R}}-\im \xi^{\text{I}}$ extends to a complexified solution in all of the solid hypercylinder, $\hat{\xi}\in L_M^{\C}$. We then have the following factorization identity for the correlation function $\rho_M^F$ of the observable $F$ in the state $\ncoh_\xi\in\cH_R$ \cite{Oe:feynobs,CoOe:locgenvac}:
\begin{equation}
  \rho_M^F(\ncoh_\xi)=\rho_M(\ncoh_\xi) F(\hat{\xi}) \rho_M^F(\ncoh_0) .
  \label{eq:corrfact}
\end{equation}
Here, $\ncoh_0$ is the vacuum state. To calculate its correlation function we need another ingredient. If we replace the action $S$ of the Klein-Gordon theory by adding the linear term $D$, the homogeneous equation of motion is replaced by an inhomogeneous equation, where we can interpret $D$ as a generalized source term. We denote the affine space of solutions in the solid hypercylinder $M$, of this inhomogeneous equation of motion by $A_M^D$ and its complexification by $A_M^{D,\C}\defeq A_M^D\oplus \im L_M$. There is then precisely one solution $\eta$ in $A_M^{D,\C}$ which satisfies the exterior vacuum boundary condition, i.e., $\eta\in A_M^{D,\C}\cap L_X^{\C}$. With this solution the vacuum correlation function is \cite{CoOe:locgenvac} 
\begin{equation}
  \rho_M^F(\ncoh_0)=\exp\left(\frac{\im}{2} D(\eta)\right) .
\end{equation}

The factorization formula (\ref{eq:corrfact}) is quite compelling in the propagating sector. In particular, observe that in the “classically allowed” case, when $\xi\in L_M^{\Lp}$ is a regular solution in the interior, the correlation function precisely recovers the classical value of the observable $F$ on the solution $\xi=\xi^{\text{R}}=\hat{\xi}$, up to the constant factor given by the vacuum correlator. In the evanescent case the situation is not as simple. Again, the first obstacle is that $\xi$ is an element of the twisted rather than the real phase space. To remedy this, we can again introduce the identification map (\ref{eq:idralpha}). The fact that this map satisfies the interior compatibility condition (\ref{eq:idintcompat}) means that we obtain a factorization identity for the correlation function rather similar to (\ref{eq:corrfact}) \cite{CoOe:locgenvac}. Defining $\hat{\tilde{\xi}}\defeq \xi^{\tilde{\text{R}}}-\im \xi^{\tilde{\text{I}}}$ we have,
\begin{equation}
  \rho_M^F(\ncoh_{I_R(\xi)})=\rho_M(\ncoh_{I_R(\xi)}) F(I_R(\hat{\tilde{\xi}})) \rho_M^F(\ncoh_0) .
\end{equation}
Even though we can now take $\xi\in L_M$, i.e., $\xi$ being a real and regular solution in the interior also in the evanescent sector, we see that in this case the observable $F$ is not evaluated on $\xi$, but on the complexified solution $I_R(\xi)$. One may take this as evidence against a correspondence between real solutions $\xi$ and coherent states $\ncoh_{I_R(\xi)}$ in the evanescent sector. However, this evidence should not be over-interpreted, because it ultimately relies on the idea of a close relationship between correlation functions and expectation values of measurements. Such a relationship is, however, quite tenuous in quantum field theory. If, instead, we think of correlators rather as instruments for introducing interactions and setting up perturbation theory, then this evidence against the correspondence between the classical solution $\xi$ and the quantum coherent state $\ncoh_{I_R(\xi)}$ becomes less relevant, while the evidence in favor in the form of the amplitude formula (\ref{eq:amplmod}) remains strong.


\section{LSZ reduction at finite distance}
\label{sec:lsz}

The LSZ reduction formula \cite{LSZ:reduction} is an essential tool in QFT as it allows expressing S-matrix elements in terms of vacuum correlation functions. A generalization of this formula to GBQFT was presented in \cite{CoOe:locgenvac}. We elaborate here on its application to the present setting. We start by recalling the standard LSZ formula as presented in textbooks \cite{ItZu:qft},
\begin{multline}
    \langle p_1,\ldots,p_m| q_1,\ldots,q_n \rangle= \text{disconnected terms}\\
    + \im^{n+m}\int \xd^4 x_1\cdots\xd^4 x_n \xd^4 y_1\cdots\xd^4 y_m
    \exp\left(\im \sum_{l=1}^m p_l\cdot y_l - \im \sum_{k=1}^n q_k\cdot x_k \right)\\
    \dop_{x_1}\cdots \dop_{x_n} \dop_{y_1}\cdots \dop_{y_m}
    \langle 0 | \tord \phi(x_1)\cdots\phi(x_n)\phi(y_1)\cdots\phi(y_m)|0\rangle .
    \label{eq:stdlsz}
\end{multline}
Here, the left-hand side represents the S-matrix between incoming particles with momenta $q_1,\ldots,q_n$ and outgoing particles with momenta $p_1,\ldots,p_m$ for Klein-Gordon theory. The notation $\dop_x$ refers to the Klein-Gordon operator $\dop\defeq \square+m^2$ as a differential operator with respect to the $x$ coordinate. The "disconnected terms" refer to terms where some of the particles do not interact. Note that in the present simplified treatment we leave out renormalization constants.

Denoting states with a notation analogous to that of Section~\ref{sec:particles} we write the left-hand side as,
\begin{equation}
    \rho_{[-\infty,\infty]}^F(\psi_{\Lin,q_1;\ldots;\Lin,q_n;\Lout,p_1;\ldots;\Lout,p_m}) .
\end{equation}
The notation $\rho_{[-\infty,\infty]}^F$ indicates the temporally asymptotic amplitude ($[-\infty,\infty]$) with interaction ($F$). In \cite{CoOe:locgenvac} we also introduced the following notation for the amplitude, where the "disconnected terms" have already been removed,\footnote{The amplitude $\rho_{\mathrm{c}}$ was also referred to as the "connected amplitude" in \cite{CoOe:locgenvac}. We do not use this term here as it might cause confusion with the notion of "connected" used in textbooks, which refers instead to connected Feynman diagrams.}
\begin{equation}
    \rho_{\mathrm{c},[-\infty,\infty]}^F(\psi_{\Lin,q_1;\ldots;\Lin,q_n;\Lout,p_1;\ldots;\Lout,p_m}) .
    \label{eq:minksmc}
\end{equation}

We proceed to recall the general procedure to obtain the LSZ formula for a spacetime region $M$ and interaction given by an observable $F$. We use the notation $\psi_{\xi_1,\ldots,\xi_n}$ with $\xi_1,\ldots,\xi_n\in L_{\partial M}^{\alpha}$ to denote a state in the space $\cH_{\partial M}$.\footnote{We suppose here the general case of an $\alpha$-Kähler quantization. In the case of ordinary Kähler quantization we set $\alpha(\phi)=\overline{\phi}$ and $L_{\partial M}^\alpha=L_{\partial M}$.} Then we have the equality \cite[(6.23)]{CoOe:locgenvac},
\begin{equation}
    \rho_{\mathrm{c},M}^F(\psi_{\xi_1,\ldots,\xi_n})=\rho_M^{E_1\cdots E_n\cdot F}(\coh_0).
    \label{eq:lszcore}
\end{equation}
Here, the right-hand side denotes the correlation function of the vacuum state for the observable $E_1\cdots E_n\cdot F$. Again, $F$ is the term inducing the interaction. The observables $E_1, \ldots, E_n$ are \emph{boundary observables} that are determined by functions $E_1',\ldots,E_n'$ on the boundary phase space $L_{\partial M}$ as follows,
\begin{equation}
    E_k'(\phi)=4\im\omega_{\partial M}(\xi_k^{\mathrm{int}},\phi) .
\end{equation}
Here we make reference to the decomposition $L_{\partial M}^{\C}=L_{M}^{\C}\oplus L_X^{\C}$ with $\xi=\xi^{\textrm{int}}+\xi^{\textrm{ext}}$. This is the direct sum decomposition of the complexified boundary phase space into the solutions that continue to the interior ($L_M^\C$) and those that satisfy the asymptotic boundary conditions in the exterior ($L_X^\C$). In the Klein-Gordon theory we can convert the boundary observable to a bulk observable as follows \cite[(6.24)]{CoOe:locgenvac},
\begin{equation}
    E'_k(\phi)=2\im\int_M \xd x\, \xi_k^{\textrm{int}}(x) (\dop\phi)(x) .
    \label{eq:bdyint}
\end{equation}
Here, $\xi_k^{\textrm{int}}$ is interpreted as a solution in $L_M^{\C}$ and $\phi$ is a field configuration in $M$. Substituting relation (\ref{eq:bdyint}) on the right-hand side in relation (\ref{eq:lszcore}) for each linear observable yields the generalized LSZ formula for Klein-Gordon theory,
\begin{equation}
    \rho_{\mathrm{c},M}^F(\psi_{\xi_1,\ldots,\xi_n})=(2 \im)^n\int_M \xd x_1\cdots\xd x_n\, \xi_1^{\mathrm{int}}(x_1)\cdots\xi_n^{\mathrm{int}}(x_n)
    \dop_{x_1}\cdots\dop_{x_n} \rho_M^{\phi(x_1)\cdots\phi(x_n)\cdot F}(\coh_0) .
    \label{eq:lszkg}
\end{equation}
Note that if $M$ is a subregion of a larger region $M'$, but the interaction encoded in $F$ (as well as any other observable, here $\phi(x_k)$) is confined to $M$, then the correlation functions are the same for both,
\begin{equation}
    \rho_M^{\phi(x_1)\cdots\phi(x_n)\cdot F}(\coh_0)=
    \rho_{M'}^{\phi(x_1)\cdots\phi(x_n)\cdot F}(\coh_0) .
\end{equation}
In particular, if $M$ is a subregion of Minkowski space taken to be $M'$, and using a traditional notation we can thus write (\ref{eq:lszkg}) as,
\begin{equation}
    \rho_{\mathrm{c},M}^F(\psi_{\xi_1,\ldots,\xi_n})=(2 \im)^n\int_M \xd x_1\cdots\xd x_n\, \xi_1^{\mathrm{int}}(x_1)\cdots\xi_n^{\mathrm{int}}(x_n)
    \dop_{x_1}\cdots\dop_{x_n}
    \langle 0 | \tord \phi(x_1)\cdots\phi(x_n)|0\rangle .
\end{equation}
If $M$ itself is Minkowski space as the asymptotic limit of a time interval region, and standard momentum states are used, we recover the standard LSZ formula, taking the form (\ref{eq:minksmc}) in traditional notation.

In the present case of interest $M$ is instead the hypercylinder $\R\times B_R^3$ in Minkowski space. As explained in Section~\ref{sec:particles}, the states are naturally parametrized by quantum numbers of energy and angular momentum. The remaining step for working out the LSZ formula of interest is thus to calculate the components $\xi^{\mathrm{int}}$ of the phase space elements $\xi$ encoding the particle states identified in Section~\ref{sec:particles}. For propagating particles we obtain with (\ref{eq:partpin}) and (\ref{eq:partpout}),
\begin{align}
    (\Phi^{\Lout,E,\ls,\ms})^{\mathrm{int}}(t,r,\Omega)
    & = \sqrt{\frac{p}{2\pi}} j_{\ls}(p r) e^{\im E t} Y_{\ls}^{-\ms}(\Omega) ,\\
    (\Phi^{\Lin,E,\ls,\ms})^{\mathrm{int}}(t,r,\Omega)
    & = \sqrt{\frac{p}{2\pi}} j_{\ls}(p r) e^{-\im E t} Y_{\ls}^{\ms}(\Omega) .
\end{align}
For the evanescent particles, we first change to a different basis that turns out to simplify the LSZ formula,
\begin{equation}
    \Phi^{\Lsi,E,\ls,\ms}=\frac{1}{\sqrt{2}}\left(\Phi^{\Lx,E,\ls,\ms}-\Phi^{\Li,E,\ls,\ms}\right), \quad \Phi^{\Lco,E,\ls,\ms}=\frac{1}{\sqrt{2}}\left(\Phi^{\Lx,E,\ls,\ms}+\Phi^{\Li,E,\ls,\ms}\right) .
\end{equation}
Then we obtain with (\ref{eq:partedec}) and (\ref{eq:parteinc}),
\begin{align}
    (\tilde{\Phi}^{\Lsi,E,\ls,\ms})^{\mathrm{int}}(t,r,\Omega)
    & = \im \sqrt{\frac{p}{2\pi^3}} a_{\ls}(p r) e^{\im E t} Y_{\ls}^{-\ms}(\Omega), \\
    (\tilde{\Phi}^{\Lco,E,\ls,\ms})^{\mathrm{int}}(t,r,\Omega)
    & = \im \sqrt{\frac{p}{2\pi^3}} a_{\ls}(p r) e^{-\im E t} Y_{\ls}^{\ms}(\Omega) .
\end{align}

Suppose the state $\psi$ on the hypercylinder consists of $n+m+k+l$ particles with quantum numbers $E_1,\ls_1,\ms_1;\ldots; E_{n+m+k+l},\ls_{n+m+k+l},\ms_{n+m+k+l}$. Additionally,
\begin{itemize}
    \item Particles $1,\ldots,n$ are \emph{incoming}.
    \item Particles $n+1,\ldots,n+m$ are \emph{outgoing}.
    \item Particles $n+m+1,\ldots,n+m+k$ are \emph{evanescent} ch-modes.
    \item Particles $n+m+k+1,\ldots,n+m+k+l$ are \emph{evanescent}  sh-modes.
\end{itemize}
With this the LSZ formula is,
\begin{multline}
    \rho_{\mathrm{c},M}^F(\psi)
    = (2\im)^{n+m+k+l} \int_{r_1,\ldots,r_{n+m+k+l}<R} \prod_{j=1}^{n+m+k+l} \xd t_j\xd\Omega_j\xd r_j r_j^2 \\
    \prod_{j=1}^n e^{-\im E_j t_j} Y_{\ls_j}^{\ms_j}(\Omega_j) \prod_{j=n+1}^{n+m} e^{\im E_j t_j} Y_{\ls_j}^{-\ms_j}(\Omega_j) \prod_{j=n+m+1}^{n+m+k} e^{-\im E_j t_j} Y_{\ls_j}^{\ms_j}(\Omega_j) \prod_{j=n+m+k+1}^{n+m+k+l} e^{\im E_j t_j} Y_{\ls_j}^{-\ms_j}(\Omega_j) \\
    \prod_{j=1}^{n+m} \sqrt{\frac{p_j}{2\pi}} j_{\ls_j}(p_j r_j)
    \prod_{j=n+m+1}^{n+m+k+l} \im\sqrt{\frac{p_j}{2\pi^3}} a_{\ls_j}(p_j r_j) \quad
    \dop_{1}\cdots \dop_{n+m+k+l} \\
    \langle 0 | \tord \phi(t_1,r_1,\Omega_1)\cdots\phi(t_{n+m+k+l},r_{n+m+k+l},\Omega_{n+m+k+l})|0\rangle .
\end{multline}
Here, $\dop_j$ denotes the Klein-Gordon operator with respect to the spacetime variables $(t_j,r_j,\Omega_j)$. Also note that the radial integrals are restricted to $0<r_j<R$.

In contrast to the original LSZ formula this exhibits the following novel features:
\begin{itemize}
    \item It applies to particles at finite distance rather than asymptotic particles.
    \item Particles are characterized in terms of energy and angular momentum quantum numbers rather than 3-momentum.
    \item It includes evanescent particles.
\end{itemize}
We observe that the calculation of the right-hand side does not require any tools or ingredients that would go beyond standard textbook QFT. In particular, the vacuum $(n+m+k+l)$-point function is the standard one on all of Minkowski space. However, recall that interactions are restricted here to occur only inside the hypercylinder of radius $R$.

In the asymptotic limit $R\to\infty$ the radial integrals run over $0<r_j<\infty$, and we may drop the restriction on the spatial restriction of the interaction. In that case, of course, the evanescent particles are not present, i.e., $k=0$ and $l=0$, shortening the formula considerably. Nevertheless, the formula still exhibits novel features compared to the standard formula (\ref{eq:stdlsz}) due to its reference to energy and angular momentum quantum numbers.


\section{Conclusions and Outlook}
\label{sec:conclude}

In the present work we have set up the building blocks of quantum scattering theory at finite distance, using the example of Klein-Gordon theory. Crucially, in doing so we have captured not only the propagating, but also the evanescent degrees of freedom of the quantum field. To this end, we have utilized the framework of general boundary quantum field theory (GBQFT) together with a recent generalization of Kähler quantization. In the following we offer a few remarks on the present work and on future directions.

For one, it is interesting to reflect on why our results are out of reach of the traditional approach to scattering in quantum field theory. As we have seen, there are at least two distinct reasons for this. The first, of a more technical nature, is that the relevant classical phase space in the standard approach consists of solutions well-behaved in all of space. Since evanescent waves are unbounded in at least one spatial direction, they are excluded from this phase space and thus from its quantization. The second reason is of a more conceptual nature. Evanescent degrees of freedom do not separate into incoming and outgoing degrees of freedom with respect to a scattering region. However, such a separation is an inherent ingredient of the standard approach based on the paradigm of transition amplitudes between incoming (initial) and outgoing (final) degrees of freedom. Of course, there is nothing wrong with the standard approach as long as we are interested in asymptotic scattering only, where the evanescent field does vanish.

In Section~\ref{sec:ampl} we have not offered further explanations of how probabilities for scattering events are computed from amplitudes like expression (\ref{eq:amplhcyl}). For the propagating sector we have seen that we can convert such amplitudes to a structure that resembles traditional transition amplitudes, compare expression (\ref{eq:inoutdef}), which may then be treated in the usual way. For the evanescent sector a mathematically analogous mapping can also be defined, but it does not admit an analogous probability interpretation as it is not based on an in/out decomposition.
However, there is a much more general and powerful way to extract probabilities from the types of amplitudes we have considered. This was first described in \cite{Oe:gbqft} and applied to scattering in the hypercylinder setting in \cite{Oe:kgtl}, to which we refer the reader. The relevant probability interpretation and formulas were later discovered to arise from first principles in the positive formalism \cite{Oe:posfound}, which we adopt here as the underlying conceptual framework.

One complication of the novel $\alpha$-Kähler quantization scheme as compared to ordinary Kähler quantization is the need for an additional structure on top of the vacuum boundary conditions of the field. This is the real structure $\alpha$, for which a priori there may be many choices. Indeed, in the quantization of the evanescent field on a timelike hyperplane, a natural 1-parameter family of such structures was found \cite{CoOe:evanescent}. It then turns out that evolving in space from one hyperplane to another one yields a unitary map between Hilbert spaces only if the same $\alpha$ is chosen for both hypersurfaces. Nevertheless, as explained in \cite{CoOe:evanescent} making different choices for different hypersurfaces is compatible with a consistent probability interpretation and does not signal physically distinguishable theories. In contrast, in the present setting of the hypercylinder we have seen that putting stringent requirements on $\alpha$ were sufficient to determine it uniquely. These were: descendence from a reflection map $\gamma$, covariance with respect to symmetries, and interior compatibility (compare Section~\ref{sec:quanthypcyl}). Consequently, as seen in Section~\ref{sec:radevol}, radial evolution is described by straightforward unitary maps.

A remaining obstacle to a fully satisfactory interpretation and application of our results concerns the correspondence between classical and quantum degrees of freedom. In Kähler quantization such a correspondence is quite clear by relating real classical observables to hermitian operators which in turn serve as ingredients of quantum measurement theory. In $\alpha$-Kähler quantization, as used here for the evanescent degrees of freedom, a correspondence is established instead between real functions on the twisted classical phase space, consisting of certain complexified solutions, and hermitian operators. To return to a desirable correspondence involving real functions on the ordinary phase space we have tentatively used a canonical identification between the ordinary and twisted phase space, as proposed in \cite{CoOe:locgenvac}. The results are encouraging in part, as we have seen in Section~\ref{sec:ampl}, but certainly cannot be considered conclusive. A dedicated further investigation of the correspondence issue is necessary.

Among the presented results is a novel LSZ reduction formula (see Section~\ref{sec:lsz}). It serves the same purpose as the original LSZ formula in providing a tool to derive scattering amplitudes from $n$-point functions. In contrast to the original formula it applies to scattering at finite distance and involves evanescent particles in addition to propagating ones. Moreover, particles are described by energy and angular momentum quantum numbers, rather than 3-momenta. However, in spite of the fact that it was derived using novel methods and ingredients, its evaluation does not require any of that and solely relies on ingredients from textbook QFT.

We have considered in the present work the Klein-Gordon theory, not because any of the methods employed is specific to this theory, but because it provides a simple and clean example. In particular, it does not introduce distractions due to gauge degrees of freedom, which are inessential for the conceptual structure of the presented results. However, for actual applications, the electromagnetic field is certainly much more interesting. In this respect it is worth mentioning that the massless Klein-Gordon theory does not actually exhibit evanescent solutions on the timelike hypercylinder. (It does on the timelike hyperplane \cite{CoOe:evanescent}). In contrast, the electromagnetic field does. Thus, our inclusion of a mass term, which does lead to evanescent solutions, can be seen as a means of mimicking this feature of the electromagnetic field.

Going further, one would eventually also want to consider the application to the gravitational field. There, one might hope to describe quantum aspects of gravitational scattering events. Of course, as is usual for methods of quantum field theory, this would work only at the level of perturbation theory. Of particular interest might be black hole spacetimes, as the phase space on a timelike hypercylinder at sufficient distance from the singularity
could be kept in a weak gravity regime. This is in contrast to any spacelike hypersurface on which a more conventional approach to quantization would be relying.

\subsection*{Acknowledgments}

I would like to thank Olivier Sarbach for inviting me to the BIRS workshop “Time-like Boundaries in General Relativistic Evolution Problems” at Oaxaca in July 2019 and giving me the opportunity to present preliminary results of this work and its underlying framework.
I would further like to thank Daniele Colosi for his longstanding collaboration, instrumental in leading to the present work.
This publication was made possible through the support of the ID\# 61466 grant from the John Templeton Foundation, as part of the “The Quantum Information Structure of Spacetime (QISS)” Project (qiss.fr). The opinions expressed in this publication are those of the author and do not necessarily reflect the views of the John Templeton Foundation.


\newcommand{\eprint}[1]{\href{https://arxiv.org/abs/#1}{#1}}
\bibliographystyle{stdnodoi} 
\bibliography{stdrefsb}
\end{document}